\documentclass[10pt,aps,twocolumn,prd,superscriptaddress,noshowpacs,nofootinbib,noshowkeys,floatfix]{revtex4}
\usepackage[dvips]{graphics,graphicx}
\usepackage[colorlinks=true,linktocpage=true,linkcolor=blue,citecolor=blue]{hyperref}
\usepackage[usenames,dvipsnames]{color}
\usepackage{amsmath, amssymb}
\usepackage{multirow}
\usepackage{longtable}
\usepackage{color}
\usepackage[normalem]{ulem}  

\renewcommand\sout{\bgroup \color{blue} \ULdepth=-.5ex \ULset}

\begin{document}

\preprint{}

\title{Relativistic second-order dissipative hydrodynamics at finite chemical potential}

\author{Amaresh Jaiswal}
\affiliation{GSI, Helmholtzzentrum f\"ur Schwerionenforschung, Planckstrasse 1, D-64291 Darmstadt, Germany}
\author{Bengt Friman}
\affiliation{GSI, Helmholtzzentrum f\"ur Schwerionenforschung, Planckstrasse 1, D-64291 Darmstadt, Germany}
\author{Krzysztof Redlich}
\affiliation{Institute of Theoretical Physics, University of Wroclaw, PL-50204 Wroclaw, Poland}
\affiliation{Extreme Matter Institute EMMI, GSI, Planckstrasse 1, D-64291 Darmstadt, Germany}

\date{\today}

\begin{abstract}

Starting from the Boltzmann equation in the relaxation time 
approximation and employing a Chapman-Enskog like expansion for the 
distribution function close to equilibrium, we derive second-order 
evolution equations for the shear stress tensor and the dissipative 
charge current for a system of massless quarks and gluons. The 
transport coefficients are obtained exactly using quantum statistics 
for the phase space distribution functions at non-zero chemical 
potential. We show that, within the relaxation time approximation, 
the second-order evolution equations for the shear stress tensor and 
the dissipative charge current can be decoupled. We find that, for 
large values of the ratio of chemical potential to temperature, the 
charge conductivity is small compared to the coefficient of shear 
viscosity. Moreover, we show that in the relaxation-time 
approximation, the limiting behaviour of the ratio of heat 
conductivity to shear viscosity is qualitatively similar to that 
obtained for a strongly coupled conformal plasma.

\end{abstract}



\maketitle

\section{Introduction}

High-energy heavy ion collisions at the BNL Relativistic Heavy Ion 
Collider (RHIC) \cite{Adams:2005dq, Adcox:2004mh} and the CERN Large 
Hadron Collider (LHC) \cite{ALICE:2011ab, ATLAS:2012at, 
Chatrchyan:2013kba} create strongly interacting matter under extreme 
conditions of high temperature and density as it is believed to have 
existed in the very early universe \cite{Rischke:2003mt, 
Tannenbaum:2006ch}. At such conditions, quarks and gluons are 
deconfined to form a new state of matter, the quark-gluon plasma 
(QGP). The QGP behaves as a strongly coupled plasma having the 
smallest shear viscosity-to-entropy density ratio, $\eta/s$ \cite 
{Romatschke:2007mq, Song:2007ux, Luzum:2009sb, Song:2010mg, 
Schenke:2010rr, Bhalerao:2015iya}. Relativistic hydrodynamics has 
been applied quite successfully to describe the space-time evolution 
of the QGP formed in high-energy heavy ion collisions and estimate 
its transport coefficients \cite{Heinz:2013th}.

In applications of hydrodynamics it is rather straightforward to 
employ the ideal (Euler) equations. The inclusion of dissipative 
effects in the evolution of the QGP started only a few years ago. 
However, most of the studies have focused on exploring the effects 
of the shear viscosity on the QGP evolution and extracting its 
magnitude from experimental measurements. Nevertheless, there are 
other sources of dissipation such as bulk viscous pressure and 
dissipative charge current that may have a significant effect on the 
hydrodynamic evolution of the QGP. While the effects of bulk viscous 
pressure has been studied in some details \cite{Monnai:2009ad, 
Denicol:2009am, Song:2009rh, Bozek:2009dw, Roy:2011pk, 
Noronha-Hostler:2013gga}, the dissipative charge current has been 
largely ignored. This may be attributed to the fact that at very 
high energies, baryon number and its corresponding chemical 
potential are negligible. However, at lower collision energies such 
as those probed in the RHIC low-energy scan or at the upcoming 
experiments at the Facility for Antiproton and Ion Research (FAIR), 
baryon number can no longer be ignored and therefore charge 
diffusion may play an important role.

The earliest theoretical formulation of relativistic dissipative 
hydrodynamics are due to Eckart \cite{Eckart:1940zz} and 
Landau-Lifshitz \cite{Landau}. However these formulations, 
collectively called relativistic Navier-Stokes theory, involve only 
first-order gradients and suffer from acausality and numerical 
instability due to the parabolic nature of the equations. Second 
order or extended theories by Grad \cite{Grad}, M\"uller \cite 
{Muller:1967zza} and Israel and Stewart (IS) \cite{Israel:1979wp} 
were introduced to restore causality. Therefore it is imperative 
that second order dissipative hydrodynamic equations should be 
employed in order to correctly describe the evolution of the QGP. 
However, the IS formulation of a causal theory of relativistic 
hydrodynamics from kinetic theory, contains several inconsistencies 
and approximations, the resolution of which is currently an active 
research area \cite{Denicol:2010xn, Jaiswal:2013fc, Jaiswal:2012qm, 
El:2009vj, Denicol:2012cn, Romatschke:2011qp, Jaiswal:2013npa, 
Jaiswal:2013vta, Bhalerao:2013aha, Bhalerao:2013pza, 
Jaiswal:2014isa, Chattopadhyay:2014lya, Florkowski:2015lra}.

In order to formulate a causal theory of relativistic dissipative 
hydrodynamics from kinetic theory, it is desirable to first specify 
the form of the non-equilibrium phase-space distribution function. 
For a system close to local thermodynamic equilibrium, the 
non-equilibrium corrections to the distribution function can be 
obtained using either (i) Grad's moment method \cite{Grad} or (ii) 
the Chapman-Enskog method \cite{Chapman}. Although both methods 
involve expanding the distribution function around its equilibrium 
value, it has been demonstrated that the Chapman-Enskog method in 
the relaxation time approximation results in a better agreement with 
microscopic Boltzmann simulations \cite{Jaiswal:2013npa, 
Jaiswal:2013vta} as well as with exact solutions of the Boltzmann 
equation in the relaxation-time approximation \cite 
{Jaiswal:2013npa, Jaiswal:2013vta, Jaiswal:2014isa, 
Chattopadhyay:2014lya, Florkowski:2015lra}. 

In the absence of conserved charges, the Chapman-Enskog method has 
been used to compute the second-order transport coefficients for 
vanishing \cite {Jaiswal:2013npa, Jaiswal:2013vta, 
Chattopadhyay:2014lya} as well as finite particle masses \cite 
{Jaiswal:2014isa, Florkowski:2015lra}. On the other hand, in the 
presence of conserved charges but for classical particles with 
vanishing masses, the second-order transport coefficients 
corresponding to charge diffusion (or alternatively heat conduction) 
have been obtained by employing the moment method \cite 
{Bouras:2010hm, Denicol:2012vq}. However, they still remain to be 
determined for quantum statistics. Here, we employ the 
Chapman-Enskog method to achieve this.

In this Letter, we present the derivation of second-order evolution 
equations for shear stress tensor and dissipative charge current for 
a system consisting of massless quarks and gluons. In order to 
obtain the form of the non-equilibrium distribution function, we 
employ a Chapman-Enskog like expansion to iteratively solve the 
Boltzmann equation in the relaxation time approximation \cite 
{Jaiswal:2013npa}. Using this expansion, we derive the first-order 
constitutive relations and subsequently the second-order evolution 
equations for the dissipative quantities. The transport coefficients 
are obtained exactly using quantum statistics for the quark and 
gluon phase-space distribution functions with a non-vanishing quark 
chemical potential. Moreover, we show that, up to second-order in 
the gradient expansion, the evolution equations for the shear stress 
tensor and the dissipative charge current can be decoupled. We also 
find that, for large values of the ratio of  chemical potential to 
temperature, the charge conductivity is small compared to the 
coefficient of shear viscosity. Finally we demonstrate that the 
limiting behaviour of the heat conductivity to shear viscosity 
ratio, obtained here in the relaxation-time approximation, is 
qualitatively identical to that of a conformal fluid in the strong 
coupling limit.


\section{Relativistic hydrodynamics}

In the case of massless partons, i.e., massless quarks and gluons, 
the conserved energy-momentum tensor and the net-quark current can 
be expressed in terms of the single particle phase-space 
distribution function as \cite{rvogt}
\begin{align}
T^{\mu\nu} &= \!\int\! dp\ p^\mu p^\nu \left[ g_q(f_q+f_{\bar q}) + g_g f_g \right]  \nonumber\\
&= \epsilon u^\mu u^\nu - P\Delta^{\mu \nu} + \pi^{\mu\nu},  \label{en_mom_ten}\\
N^\mu &= \!\int\! dp\ p^\mu \left[ g_q(f_q-f_{\bar q}) \right] = nu^\mu + n^\mu, \label{cons_curr}
\end{align}
where $dp = d{\bf p}/[(2 \pi)^3|{\bf p}|]$, $p^\mu$ is the particle 
four momenta, and $g_q$ and $g_g$ are the quark  and gluon 
degeneracy factor, respectively. Here $f_q$, $f_{\bar q}$, and $f_g$ 
are the phase-space distribution functions for quarks, anti-quarks, 
and gluons. In the tensor decompositions, $\epsilon$, $P$, and $n$ 
are the energy density, pressure, and the net quark number density. 
The projection operator $\Delta^{\mu\nu}=g^{\mu\nu}-u^\mu u^\nu$ is 
orthogonal to the hydrodynamic four-velocity $u^\mu$ defined in the 
Landau frame: $T^{\mu\nu}u_\nu=\epsilon u^\mu$. We work with the 
Minkowskian metric tensor $g^{\mu\nu}\equiv\mathrm{diag}(+,-,-,-)$.

The dissipative quantities in Eqs.~(\ref{en_mom_ten}) and (\ref 
{cons_curr}) are the shear stress tensor $\pi^{\mu\nu}$ and the 
particle diffusion current $n^\mu$. With the definition of the 
energy-momentum tensor in Eq.~(\ref {en_mom_ten}), the bulk viscous 
pressure vanishes in the massless case. The energy-momentum 
conservation, $\partial_\mu T^{\mu\nu} =0$, and particle 
four-current conservation, $\partial_\mu N^{\mu}=0$, yields the 
fundamental evolution equations for $\epsilon$, $u^\mu$ and $n$, as
\begin{align}
\dot\epsilon + (\epsilon+P)\theta - \pi^{\mu\nu}\sigma_{\mu\nu} &= 0,  \label{evol1}\\
(\epsilon+P)\dot u^\alpha - \nabla^\alpha P + \Delta^\alpha_\nu \partial_\mu \pi^{\mu\nu}  &= 0, \label{evol2}\\
\dot n + n\theta + \partial_\mu n^{\mu} &=0. \label{evol3}
\end{align}
Here we use the standard notation $\dot A=u^\mu\partial_\mu A$ for
co-moving derivatives, $\theta\equiv\partial_\mu u^\mu$ for the
expansion scalar, $\sigma^{\mu\nu}\equiv\frac{1}{2}(\nabla^\mu
u^\nu+\nabla^\nu u^\mu)-\frac{1}{3}\theta\Delta^{\mu\nu}$ for the
velocity stress tensor, and $\nabla^\alpha=\Delta^{\mu\alpha}
\partial_\mu$ for space-like derivatives.

In the following, we briefly outline the thermodynamic properties of 
a QGP in equilibrium. In this case, the phase-space distribution 
functions for quarks, anti-quarks and gluons are given by
\begin{align}
f^{(0)}_q &= \frac{1}{\exp(\beta\,u\cdot p-\alpha)+1}, \label{eq_dist_q}\\
f^{(0)}_{\bar q} &= \frac{1}{\exp(\beta\,u\cdot p+\alpha)+1}, \label{eq_dist_qb}\\
f^{(0)}_g &= \frac{1}{\exp(\beta\,u\cdot p)-1}, \label{eq_dist_g}
\end{align}
respectively, where $u\cdot p\equiv u_\mu p^\mu$, $\beta=1/T$ is the 
inverse temperature and $\alpha=\mu/T$ is the ratio of the quark 
chemical potential to temperature. We consider vanishing chemical 
potential for gluons because they are unconstrained by the 
conservation laws.

The temperature, $T$, and chemical potential, $\mu$, of the system 
is determined by the matching condition $\epsilon=\epsilon_0$ and 
$n=n_0$, where $\epsilon_0$ and $n_0$ is the energy density and the 
net quark number density in equilibrium. The energy density, 
pressure and the net quark number density for a system of massless 
quarks and gluons in equilibrium is given by
\begin{align}
\epsilon_0 &\equiv u_\mu u_\nu \!\int\! dp \, p^\mu p^\nu \left[ g_q\left(f^{(0)}_q + f^{(0)}_{\bar q}\right)
+ g_g f^{(0)}_g \right] \nonumber\\
&= \frac{(4g_g+7g_q)\pi^2}{120}T^4 + \frac{g_q}{4}T^2\mu^2 + \frac{g_q}{8\pi^2}\mu^4 \label{en_den}\\
P_0 &\equiv -\frac{1}{3}\Delta_{\mu\nu} \!\int\! dp \, p^\mu p^\nu \left[ g_q\left(f^{(0)}_q + f^{(0)}_{\bar q}\right)
+ g_g f^{(0)}_g \right] \nonumber\\
&= \frac{(4g_g+7g_q)\pi^2}{360}T^4 + \frac{g_q}{12}T^2\mu^2 + \frac{g_q}{24\pi^2}\mu^4 \label{prs} \\
n_0 &\equiv u_\mu \!\int\! dp \, p^\mu \left[ g_q\left(f^{(0)}_q - f^{(0)}_{\bar q}\right) \right] \nonumber\\
&= \frac{g_q}{6}T^2\mu + \frac{g_q}{6\pi^2}\mu^3. \label{num_den}
\end{align}
The equilibrium entropy density then becomes
\begin{equation}\label{ent_den}
s_0\equiv \frac{\epsilon_0+P_0-\mu\,n_0}{T} = \frac{(4g_g+7g_q)\pi^2}{90}T^3 + \frac{g_q}{6}T\mu^2.
\end{equation}

The above expressions for $\epsilon_0$, $P_0$, $n_0$, and $s_0$ can 
also be obtained directly from the partition function of an ideal 
QGP \cite{rvogt},
\begin{equation}\label{part_fn}
\ln Z = \frac{V}{T}\left[\frac{(4g_g+7g_q)\pi^2}{360}T^4 + \frac{g_q}{12}T^2\mu^2 + \frac{g_q}{24\pi^2}\mu^4\right],
\end{equation}
where $V$ is the volume of the system. Indeed, using the 
thermodynamic relations
\begin{align}\label{th_rels}
\epsilon_0 &= \frac{T^2}{V}\frac{\partial\ln Z}{\partial T} + \mu n_0 ,\quad
P_0 = \frac{T}{V}\ln Z, \nonumber\\
n_0 &= \frac{T}{V}\frac{\partial\ln Z}{\partial\mu} ,\quad
s_0 = \frac{1}{V}\frac{\partial(T\ln Z)}{\partial T},
\end{align}
one recovers Eqs.~(\ref{en_den})-(\ref{ent_den}). The matching 
conditions $\epsilon=\epsilon_0$ and $n=n_0$ allows us to define 
thermodynamic quantities like temperature and chemical potential of 
a dissipative system. The pressure $P$ can then be obtained from the 
equation of state of the system.

Even if the equation of state, relating $\epsilon$, $P$ and $n$ is 
provided, Eqs.~(\ref{evol1})-(\ref{evol3}) are not closed unless the 
dissipative quantities $\pi^{\mu\nu}$ and $n^\mu$ are specified. 
However, before we derive the evolution equations for the 
dissipative quantities, we need to obtain expressions for the 
derivatives of $\beta$ and $\alpha$. Using Eqs.~(\ref{evol1})-(\ref 
{evol3}) and Eqs.~(\ref {en_den})-(\ref {num_den}), we get
\begin{align}
\dot\beta &= \frac{\beta}{3}\theta + {\cal O}(\delta^2), \quad \dot\alpha = {\cal O}(\delta^2), \label{ab_dot}\\
\nabla^\mu\beta &= - \beta\dot u^\mu + \frac{n}{\epsilon+P}\nabla^\mu\alpha
- \frac{\beta}{\epsilon+P}\Delta^\mu_\rho\partial_\gamma\pi^{\rho\gamma}, \label{derv_b}
\end{align}
where ${\cal O}(\delta^2)$ represents terms which are of second or 
higher order in derivatives. Since dissipative forces are caused by 
thermodynamic gradients present in a non-ideal system, $\pi^{\mu\nu}$
and $n^\mu$ are at least linear in the gradient expansion. Note 
that while Eq.~(\ref{ab_dot}) is terminated at first-order 
(sufficient for the present work), Eq.~(\ref{derv_b}) is exact.

The QGP is a strongly coupled system and is conjectured to be formed 
close to local thermodynamic equilibrium. Therefore, the phase-space 
distribution function can be split into equilibrium and 
non-equilibrium parts, $f=f^{(0)}+\delta f$, where $|\delta 
f|/f^{(0)}\ll 1$. Hence, from Eqs.~(\ref{en_mom_ten}) and (\ref 
{cons_curr}), the shear stress tensor $\pi^{\mu\nu}$ and the 
particle diffusion current $n^\mu$ can be expressed in terms of 
$\delta f$ as
\begin{align}
\pi^{\mu\nu} &= \Delta^{\mu\nu}_{\alpha\beta} \!\int\! dp \, p^\alpha p^\beta
\left[g_q(\delta f_q + \delta f_{\bar q}) + g_g\delta f_g\right] ,\label{shear}\\
n^\mu &= \Delta^\mu_\alpha \!\int\! dp \, p^\alpha
\left[g_q(\delta f_q - \delta f_{\bar q})\right] , \label{ch_curr}
\end{align}
where $\Delta^{\mu\nu}_{\alpha\beta}\equiv \frac{1}{2} 
(\Delta^{\mu}_{\alpha}\Delta^{\nu}_{\beta} + 
\Delta^{\mu}_{\beta}\Delta^{\nu}_{\alpha}) - \frac{1}{3} 
\Delta^{\mu\nu}\Delta_{\alpha\beta}$ is a traceless symmetric 
projection operator orthogonal to $u_\mu$ and $\Delta_{\mu\nu}$. In 
the following, we obtain $\delta f$ up to first order by using the 
iterative solution of the Boltzmann equation in the relaxation-time 
approximation and then derive second-order evolution equations for 
the dissipative quantities.


\section{Dissipative evolution equations}

The determination of the form of the non-equilibrium phase-space 
distribution function is a central problem in statistical physics. 
This can be achieved by solving a kinetic equation like the 
Boltzmann equation. The Boltzmann equation governs the evolution of 
the distribution function which provides a complete description of 
the microscopic dynamics of a system in the dilute limit. The 
relativistic Boltzmann equation with the collision term written in 
the relaxation-time approximation is \cite 
{Anderson_Witting},
\begin{equation}\label{boltz_eq}
p^\mu\partial_\mu f =  -\frac{u\cdot p}{\tau_R}\left( f-f^{(0)} \right),
\end{equation}
where $\tau_R$ is the relaxation time. Note, that for different 
species of particles, with inter- and intra-species interactions, 
the relaxation-times are usually distinct. Thus, in general, one 
should consider the QGP as a true multicomponent system. In the 
following, we consider the special case with a common relaxation 
time for all particle species in the QGP. The general case with a 
different relaxation time is left for future work.

We employ iterative solution of the Boltzmann equation (\ref
{boltz_eq}) to derive the dissipative equations \cite 
{Romatschke:2011qp, Jaiswal:2013npa, Jaiswal:2013vta}. The 
first-order expressions for shear stress tensor and dissipative 
charge current are obtained as \cite{Jaiswal:2013npa},
\begin{equation}\label{first_ord_diss}
\pi^{\mu\nu} = 2\beta_\pi\tau_R\sigma^{\mu\nu}, \quad
n^\mu = \beta_n\tau_R\nabla^\mu\alpha.
\end{equation}
Here $\beta_\pi$ and $\beta_n$ are the first-order transport 
coefficients obtained after performing the momentum integrations in 
Eqs.~(\ref{shear}) and (\ref{ch_curr}). For a system of massless 
partons, as considered here, we find
\begin{equation}\label{beta_pi_n}
\beta_\pi = \frac{\epsilon + P}{5}, \quad
\beta_n = \frac{J_{10}^+}{3} - \frac{n^2\,T}{\epsilon+P},
\end{equation}
where,
\begin{align}\label{J10p}
J_{10}^+ \,&\, \equiv g_q\!\int\!dp\,(u\cdot p)\left( f^{(0)}_q\tilde f^{(0)}_q
+ f^{(0)}_{\bar q}\tilde f^{(0)}_{\bar q} \right) \nonumber\\
\,&\, = \frac{g_q}{6}T^3 + \frac{g_q}{2\pi^2}T\mu^2
\,=\,\frac{\pi^2+3\alpha^2}{\alpha\left(\pi^2+\alpha^2\right)}\,n.
\end{align}
Here $\tilde f^{(0)}\equiv1-rf^{(0)}$, where $r=1$ for Fermions
(quarks and anti-quarks) and $r=-1$ for Bosons (gluons).

Using Eq.~(\ref{first_ord_diss}), we also obtain the first-order 
dissipative corrections to the distribution function,
\begin{align}
\frac{\delta f^{(1)}_q}{f^{(0)}_q\!\tilde f^{(0)}_q} &\!= \frac{\beta}{2(u\!\cdot\!p)\beta_\pi}p^\mu p^\nu \pi_{\mu\nu}
+ \frac{1}{\beta_n}\!\!\left(\! \frac{n}{\epsilon\!+\!P}
- \frac{1}{u\!\cdot\!p} \!\right)\!p^\mu n_\mu , \label{neq_dist_q}\\
\frac{\delta f^{(1)}_{\bar q}}{f^{(0)}_{\bar q}\!\tilde f^{(0)}_{\bar q}} &\!= \frac{\beta}{2(u\!\cdot\!p)\beta_\pi}p^\mu p^\nu \pi_{\mu\nu}
+ \frac{1}{\beta_n}\!\!\left(\! \frac{n}{\epsilon\!+\!P}
+ \frac{1}{u\!\cdot\!p} \!\right)\!p^\mu n_\mu , \label{neq_dist_qb}\\
\frac{\delta f^{(1)}_g}{f^{(0)}_g\!\tilde f^{(0)}_g} &\!= \frac{\beta}{2(u\!\cdot\!p)\beta_\pi}p^\mu p^\nu \pi_{\mu\nu}. \label{neq_dist_g}
\end{align}
Note, that the tensorial form of dissipative corrections to the
distribution function, as given in the above equations, is analogous
to that of Grad's 14-moment approximation \cite{Grad}. However, the
coefficients of these terms are different which has interesting
implications in the context of relativistic heavy-ion collisions
\cite{Bhalerao:2013pza}.

To obtain the second-order evolution equations, we follow the 
procedure discussed in Ref.~\cite{Denicol:2010xn}. We consider the 
comoving derivative of Eqs.~(\ref{shear}) and (\ref{ch_curr}), and 
rewrite Eq.~(\ref{boltz_eq}) in favour of $\delta\dot f$. Using 
Eqs.~(\ref{neq_dist_q})-(\ref{neq_dist_g}) and performing the 
momentum integrations, we finally obtain the second-order evolution 
equation for $\pi^{\mu\nu}$ and $n^\mu$,
\begin{align}
\dot\pi^{\langle\mu\nu\rangle} + \frac{\pi^{\mu\nu}}{\tau_\pi} =&~
2\beta_\pi\sigma^{\mu\nu}
+ 2\pi_\gamma^{\langle\mu}\omega^{\nu\rangle\gamma}
-\frac{4}{3}\pi^{\mu\nu}\theta \nonumber \\
&-\frac{10}{7}\pi_\gamma^{\langle\mu}\sigma^{\nu\rangle\gamma}, \label{sh_final}\\
\dot n^{\langle\mu\rangle} + \frac{n^\mu}{\tau_n} =&~
\beta_n\nabla^\mu\alpha
-n_\nu\omega^{\nu\mu}
-n^\mu\theta
-\frac{3}{5}n_\nu\sigma^{\nu\mu} \nonumber \\
&-\frac{3\,\beta_n}{\epsilon+P}\,\pi^{\mu\nu}\nabla_\nu\alpha. \label{ch_sem_final}
\end{align}
Here $\omega^{\mu\nu}\equiv(\nabla^\mu u^\nu-\nabla^\nu u^\mu)/2$ is 
the anti-symmetric vorticity tensor and we have ignored terms higher 
than quadratic order in the gradients \cite{Jaiswal:2013npa}. Note 
that in the relaxation-time approximation, the Boltzmann relaxation 
time $\tau_R$ is the time scale for evolution of both $\pi^{\mu\nu}$ 
and $n^\mu$, i.e., $\tau_\pi=\tau_n=\tau_R$. By comparing the 
first-order equations, Eq.~(\ref{first_ord_diss}), with the 
relativistic Navier-Stokes equations for the dissipative quantities 
\cite{Landau},
\begin{equation}\label{rel_NS_eqn}
\pi^{\mu\nu} = 2\eta\sigma^{\mu\nu}, \quad
n^\mu = \kappa_n\nabla^\mu \alpha,
\end{equation}
the dissipative relaxation times can be related to the first-order 
transport coefficients $\tau_\pi=\eta/\beta_\pi$ and 
$\tau_n=\kappa_n/\beta_n$.

It is interesting to note that terms like $\Delta^\mu_\nu 
\partial_\gamma \pi^{\nu\gamma}$ and $\pi^{\mu\nu}\dot u_\nu$ do not 
appear in Eq.~(\ref{ch_sem_final}) because their coefficients 
vanish. Note also that, with $\beta_\pi=(\epsilon+P)/5$ calculated 
using the appropriate equation of state, Eq.~(\ref{sh_final}) is 
valid even in a hadronic phase dominated by massless pions.

The last term in Eq.~(\ref{ch_sem_final}) couples the evolution of 
dissipative charge current with the shear stress tensor. This type 
of coupling leads to disagreement with transport results as shown in 
Ref.~\cite{Bouras:2010hm}. We observe, however, that using Eq.~(\ref 
{first_ord_diss}), the last term in Eq.~(\ref {ch_sem_final}) is, up 
to second-order in the gradient expansion, equivalent to 
$-(6/5)n_\nu\sigma^{\nu\mu}$. Thus, the evolution equation, Eq.~(\ref
{ch_sem_final}), for the dissipative charge current becomes
\begin{equation}\label{ch_final}
\dot n^{\langle\mu\rangle} + \frac{n^\mu}{\tau_n} =
\beta_n\nabla^\mu\alpha - n_\nu\omega^{\nu\mu}
- n^\mu\theta - \frac{9}{5}n_\nu\sigma^{\nu\mu}.
\end{equation}
This is the main result of the present work. The compact form of the 
above equation makes it straightforward for direct implementation in 
a viscous hydrodynamic code. Note that while it is possible to 
formally rewrite any second-order equation only in terms of 
gradients using the first-order expressions, Eq.~(\ref 
{first_ord_diss}), it does not usually imply decoupling of the 
dissipative evolution equations \cite {Finazzo:2014cna}. However, we 
have ensured that the second-order terms in Eq.~(\ref {ch_final}) is 
product of a dissipative quantity and a gradient, as inherent in 
Eq.~(\ref {sh_final}).

It is important to note that the gradient expansion converges only 
for small deviations from equilibrium. This implies that the 
second-order scheme is, strictly speaking, justified only if the 
deviations from the first-order relations, Eq.~(\ref 
{first_ord_diss}), are of second order, or smaller. This is true, in 
general, to ensure the convergence of gradient expansion in the 
formulation of dissipative hydrodynamics. Hence, for consistency, 
the initial conditions for the second-order evolution equations, 
Eqs.~(\ref{sh_final}) and (\ref {ch_final}), should be chosen such 
that the constitutive relations, Eq.~(\ref {first_ord_diss}), are 
satisfied. Nevertheless, it was shown in Refs.~\cite 
{Jaiswal:2013vta, Jaiswal:2014isa, Chattopadhyay:2014lya, 
Florkowski:2015lra} that the solutions of the second-order 
dissipative hydrodynamic equations rapidly converge to the exact 
solution of the Boltzmann equation, irrespective of the choices of 
initial conditions for the dissipative quantities. Thus, the use of 
Eq.~(\ref {first_ord_diss}), to modify the second-order terms in the 
evolution equation for the dissipative charge current, is tenable 
even if the initial conditions are chosen such that these relations 
are violated initially. 


\section{Transport coefficients}

The dimensionless ratio $\kappa_n T/\eta$ is a measure of the 
relative importance of the charge conductivity and the shear 
viscosity. This quantity, in the relaxation-time approximation, is 
given by $\kappa_n T/\eta=\beta_n T/\beta_\pi$, which can be studied 
as a function $\mu/T$. To quantify this ratio, one still need to 
specify the appropriate quark and gluon degeneracy factors, $g_q$ 
and $g_g$, as
\begin{align}\label{degen}
g_q &= N_s \times N_c \times N_f = 6\,N_f, \nonumber\\
g_g &= N_s \times \big(N_c^2 - 1\big) = 16,
\end{align}
where $N_s=2$ is the number of spin degrees of freedom, $N_c=3$ is 
the number of colours, and $N_f$ is the number of flavours.

\begin{figure}[t]
\begin{center}
\includegraphics[scale=0.4]{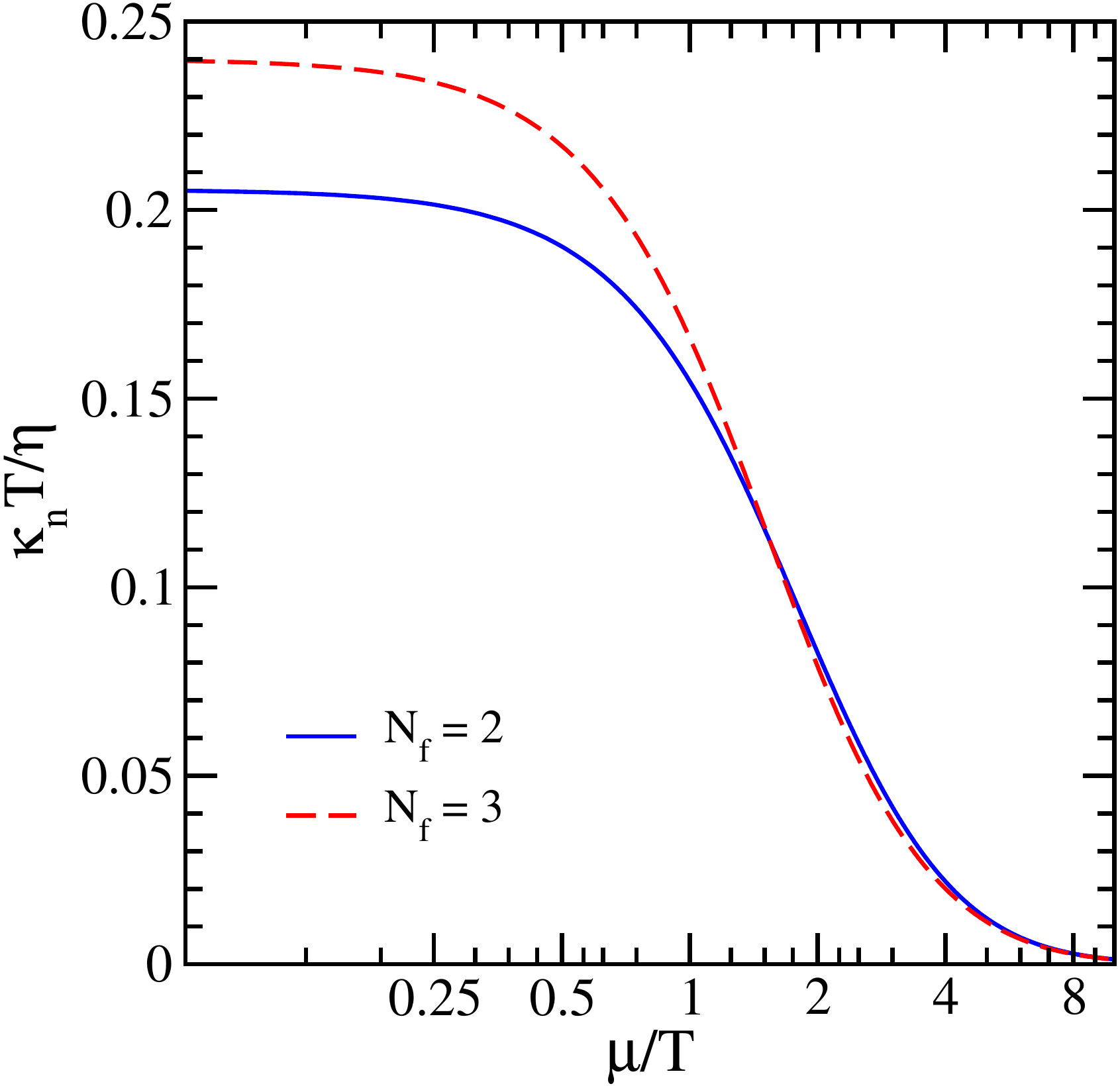}
\end{center}
\vspace{-0.5cm}
\caption{(Color online) The ratio of charge conductivity to shear 
viscosity scaled by the temperature, $\kappa_n T/\eta$, for two 
flavor (solid line) and three flavor (dashed line) massless  quarks, 
plotted against $\mu/T$.}
\label{kappanbyeta}
\end{figure}

In Fig.~\ref{kappanbyeta}, we show the ratio $\kappa_n T/\eta$ as a 
function of $\mu/T$ for $N_f=2$ and $N_f=3$. We observe that, while 
for small $\mu/T$, this ratio is almost constant, it drops rapidly 
for larger $\mu/T$, indicating that the conductivity of the QGP is 
small relative to the shear viscosity at low temperature and high 
density. Although the qualitative behaviour remains the same for 
$N_f=2$ and $N_f=3$, the drop in $\kappa_n T/\eta$ is more 
pronounced for $N_f=3$. Moreover, we note that at $\mu/T>1$ the 
ratio $\kappa_n T/\eta$ is almost $N_f$ independent.

\begin{figure}[t]
\begin{center}
\includegraphics[scale=0.4]{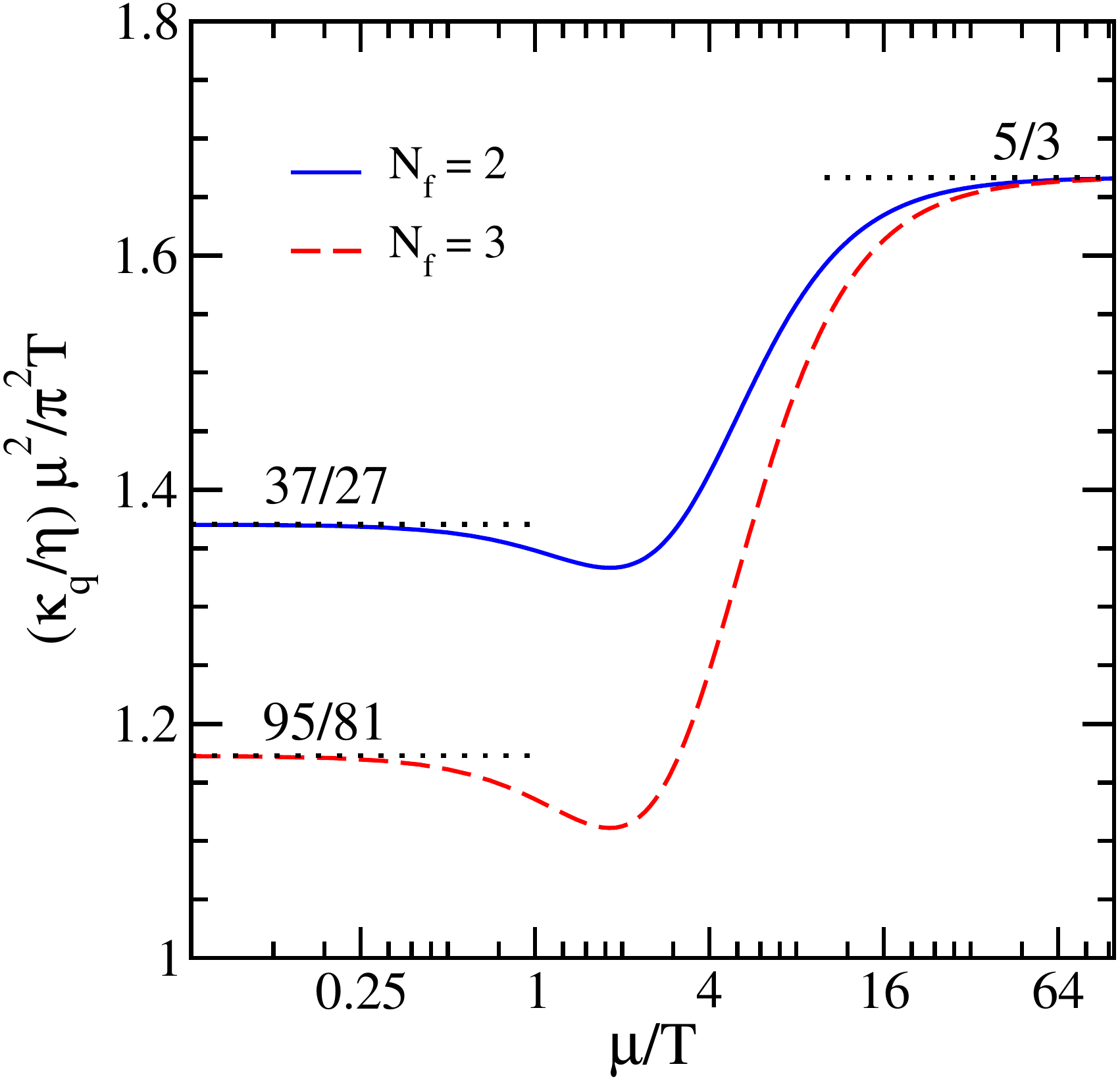}
\end{center}
\vspace{-0.5cm}
\caption{(Color online) The ratio of thermal conductivity to shear 
viscosity, $\kappa_q/\eta$, scaled by the factor $\mu^2/\pi^2T$, for 
two flavor (solid line) and three flavor (dashed line) massless 
quarks, plotted against $\mu/T$.}
\label{kappaqbyeta}
\end{figure}

A further interesting quantity is the ratio of thermal conductivity 
to shear viscosity. The heat flow is related to the dissipative 
charge current via the relation $q^\mu=-(\epsilon+P) n^\mu/n$, and 
is given as \cite{Son:2006em}
\begin{equation}\label{heat_flow}
q^\mu = -\kappa_q\,\frac{n\,T^2}{\epsilon+P}\,\nabla^\mu\alpha
\quad\Rightarrow\quad
\kappa_q = \kappa_n\left(\frac{\epsilon+P}{n\,T}\right)^2,
\end{equation}
where $\kappa_q$ is the coefficient of thermal conductivity. Using 
the first-order relations in the relaxation-time approximation, we 
find
\begin{equation}\label{conduct_visc}
\frac{\kappa_q}{\eta} = \frac{\beta_n}{\beta_\pi}\left(\frac{\epsilon+P}{n\,T}\right)^2.
\end{equation}
In Fig.~\ref{kappaqbyeta}, we plot $\kappa_q/\eta$ scaled by the 
factor $\mu^2/\pi^2T$ versus $\mu/T$, for a two and three flavor 
QGP. We observe a constant behaviour in the limit of small as well 
as large $\mu/T$. Moreover, for large $\mu/T$, we see that 
$(\kappa_q/\eta)\mu^2/\pi^2T$ is independent of number of flavors. 
These limiting behaviours have interesting consequences.

In the limit of both small and large $\mu/T$, Eq.~(\ref 
{conduct_visc}) reduces to
\begin{equation}\label{conduct_visc_lim}
\frac{\kappa_q}{\eta} = C\,\frac{\pi^2\,T}{\mu^2},
\end{equation}
which is similar to the Wiedemann-Franz law \cite 
{Pitaevskii_Lifshitz, Son:2006em}. The factor $\pi^2$ in the above 
equation is due to quantum statistics and it does not appear for a 
classical Boltzmann gas. In the limit of small $\mu/T$, the constant 
$C$ in Eq.~(\ref{conduct_visc_lim}) is $C=(4g_g+7g_q)/9g_q$. Thus, 
$C=37/27$ and $95/81$ for two and three flavor QGP, respectively; 
see Fig.~\ref{kappaqbyeta}. On the other hand, for large $\mu/T$ we 
find $C=5/3$, independent of the number of flavors, as shown in 
Fig.~\ref{kappaqbyeta}. These values of $C$ are comparable to 
$C=8/9$ (here a factor of $1/9$ indicates that the baryon chemical 
potential is three times the quark chemical potential) obtained in 
the calculations for strongly coupled conformal plasmas with finite 
chemical potential \cite{Son:2006em}. However, it should be noted 
that for a strongly coupled conformal plasma, the coefficient $C$ 
depends on the number of space-time dimensions and is shown to be 
equal to $32/9$, $8/9$ and $2/9$ for four, five and seven 
dimensions, respectively \cite{Jain:2009pw}. Nevertheless, it is 
intriguing that up to a constant of proportionality, the limiting 
behaviour of the ratio $\kappa_q/\eta$, obtained in the 
relaxation-time approximation, is exactly the same as that derived 
in the case of a strongly coupled conformal fluid.


\section{Conclusions and outlook}

In this paper we employed the iterative Chapman-Enskog method to 
derive the second-order dissipative hydrodynamical equations for a 
system of massless quarks and gluons. The bulk viscous pressure 
vanishes for such a system and therefore the dissipation is solely 
due to the shear stress tensor and the dissipative charge current. 
For the equilibrium distribution function, we considered quantum 
statistics with non vanishing quark chemical potential. We obtained 
novel, exact relations for the second-order transport coefficients 
corresponding to the dissipative charge current evolution. 

Moreover, we demonstrated that the evolution equations for shear 
stress tensor and dissipative charge current can be decoupled. We 
also found that, for large values of the ratio of chemical potential 
to temperature, the charge conductivity is small relative to the 
shear viscosity. Finally, we showed that in the relaxation-time 
approximation, the limiting behaviour of the ratio of heat 
conductivity to shear viscosity is qualitatively similar to that of 
a conformal fluid in the strong coupling regime.

At this juncture, we would like to stress that the iterative 
Chapman-Enskog approach employed here to obtain the dissipative 
evolution equations from the Boltzmann equation in the 
relaxation-time approximation is compatible with the gradient 
expansion inherent in the formulation of dissipative hydrodynamics, 
as opposed to the moment method \cite{Jaiswal:2013npa}. Looking 
forward, it would be interesting to determine the effect of the 
dissipative charge current in high-energy heavy-ion collisions, by 
implementing the dissipative equations derived here, in realistic 
hydrodynamic simulations. A further challenging problem would be to 
extend the current second-order formulation to third order. We leave 
these questions for future studies.

\begin{acknowledgments}
A.J. thanks Gabriel Denicol for useful discussions. A.J. was 
supported by the Frankfurt Institute for Advanced Studies (FIAS). 
The work of B.F. was supported in part by the Extreme Matter 
Institute EMMI. K.R. acknowledges support by the Polish Science 
Foundation (NCN), under Maestro grant DEC-2013/10/A/ST2/00106. 
\end{acknowledgments}

\end{document}